\documentclass[journal=apchd5,manuscript=article]{achemso}

\usepackage[version=3]{mhchem} % Formula subscripts using \ce{}
\usepackage[utf8]{inputenc}
\usepackage[english]{babel}
\usepackage{amsmath,amsfonts,amsthm,amssymb}
\usepackage{mathtools}
\usepackage{graphicx}
\usepackage{color}
\usepackage[pdftex]{hyperref}
\usepackage{bm}
\usepackage[inline]{enumitem}
\usepackage{lipsum}
\usepackage{float}
\usepackage{array}
\usepackage{verbatim}
\usepackage{microtype}

\newcommand{\ep}{\epsilon}

\author{P.~A.~D.~Gon\c{c}alves}
\affiliation[]{Department of Photonics Engineering, Technical University of Denmark, DK-2800 Kgs.~Lyngby, Denmark}
\alsoaffiliation[]{Center for Nano Optics, University of Southern Denmark, Campusvej 55, DK-5230 Odense~M, Denmark}
\alsoaffiliation[]{Center for Nanostructured Graphene, Technical University of Denmark, DK-2800 Kgs.~Lyngby, Denmark}
\email{padgo@fotonik.dtu.dk}
\author{Sanshui~Xiao}
\affiliation[]{Department of Photonics Engineering, Technical University of Denmark, DK-2800 Kgs.~Lyngby, Denmark}
\alsoaffiliation[]{Center for Nanostructured Graphene, Technical University of Denmark, DK-2800 Kgs.~Lyngby, Denmark}
\author{N.~M.~R.~Peres}
\affiliation[]{Department of Physics and Center of Physics, and QuantaLab, University of Minho, PT-4710-057, Braga, Portugal}
\author{N.~Asger~Mortensen}
\affiliation[]{Center for Nano Optics, University of Southern Denmark, Campusvej 55, DK-5230 Odense~M, Denmark}
\alsoaffiliation[]{Danish Institute for Advanced Study, University of Southern Denmark, Campusvej 55, DK-5230 Odense~M, Denmark}
\alsoaffiliation[]{Center for Nanostructured Graphene, Technical University of Denmark, DK-2800 Kgs.~Lyngby, Denmark}
\email{asger@mailaps.org}

\title[Hybridized Plasmons in 2D Nano-slits]{Hybridized Plasmons in 2D Nano-slits: From Graphene to Anisotropic 2D Materials}


\begin{document}

% -------------------------------------------------------------
% #                          Abstract                         #
% -------------------------------------------------------------

\begin{abstract}
Plasmon coupling and hybridization in complex nanostructures constitutes a fertile playground for 
controlling light at the nanoscale. Here, we present a semi-analytical model to describe the emergence 
of hybrid plasmon modes guided along 2D nano-slits. In particular, we find two new coupled plasmonic 
resonances arising from symmetric and antisymmetric hybridizations of the edge plasmons of the constituent 
half-sheets. These give rise to an antibonding and a bonding mode, lying above and below the energy of the 
bare edge plasmon. Our treatment is notably generic, being able to account for slits of arbitrary width, 
and remains valid irrespective of the 2D conductive material (e.g., doped graphene, 
2D transition metal dichalcogenides, or phosphorene). We derive the dispersion relation 
of the hybrid modes of a 2D nano-slit along with the corresponding induced potential 
and electric field distributions. We also discuss the plasmonic spectrum of a 2D slit 
together with the one from its complementarity structure, that is, a ribbon. 
Finally, the case of a nano-slit made from an anisotropic 2D material is considered. Focusing 
on black phosphorus (which is highly anisotropic), we investigate the features of its plasmonic spectrum 
along the two main crystal axes.
Our results offer insights into the interaction of plasmons in complex 2D nanostructures, 
thereby expanding the current toolkit of plasmonic resonances in 2D materials, 
and paving the way for the emergence of future compact devices based on atomically thin plasmonics.
\end{abstract}

\bigskip
\bigskip
\bigskip
% *************************************************************
% ::                        INTRODUCTION                     ::
% *************************************************************
%\section{Introduction}

Surface plasmons, collective oscillations of the electronic density in conductors, 
have been under the spotlight of the nanophotonics community over the last decade\cite{Atwater} owing to their ability to confine optical fields 
below the diffraction limit.\cite{Gramotnev:2010,Nat424} As of today, the annual number of plasmon-related publications is approaching a five-digit figure.\cite{next10} 
Moreover, with the emergence of new opportunities on the horizon---such as plasmons in layered two-dimensional (2D) materials\cite{review2Dphoton,Abajo_vdWreview,low2016polaritons} 
and quantum plasmonics\cite{nphysQP,Asger_quantum}---the interest in the subject is likely to remain elevated over the next decade.\cite{next10}

The potential surface plasmons have to squeeze light into subwavevelength regimes leads to large electric field enhancements, which can be of several orders of magnitude, 
thereby promoting strong light-matter interactions in nanoscale environments. On the other hand, the local increase in the electric field strength near plasmonic nanostructures makes them 
particularly well-suited for biochemical sensing and surface-enhanced Raman scattering (SERS) applications.\cite{nphot6,nl14,Rodrigo10072015,mrs30} 
Other ensuing phenomena includes the manifestation of large nano-optical forces,\cite{C4FD00224E,Novotny} enhanced nonlinearities,\cite{NL_plasm,NL_plasmSpec} 
and modification of the spontaneous emission rate of quantum emitters.\cite{PhysRevLett.96.113002,nphysQP,QE_Pors}

It is well known that the degree of field localization attained by surface plasmons is substantially higher within small gaps between metallic structures, e.g., 
inside metal grooves,\cite{AsgerCPP_rev,SorenEELS} plasmonic dimers and bow-ties,\cite{Zohar201426,jpcl} or in the so-called particle on a mirror geometry.\cite{subnm_gaps,jjbau17} 
Indeed, in the latter configuration, sub-nanometric gaps with extremely small modal volumes have been reported using either 
self-assembled molecular monolayers\cite{subnm_gaps2,Benz726} or atomically thin 2D crystals.\cite{ACS_atomthick,subnm_G} However, 
and despite its many alluring properties, three-dimensional (3D) metal-based plasmonics currently faces two main obstacles which 
have been hampering its natural evolution from research laboratories to everyday technological devices relying on surface plasmons. The first, and often the most important, 
is unarguably the relatively short life-time of plasmons, since metals are inherently lossy. The other is associated with the limited tunability of the 
plasmon resonance, that tends to remain essentially fixed for a given geometry and material. In this panorama, graphene and other 2D materials 
beyond graphene (for instance, doped transition metal dichalcogenides (TMDCs), black phosphorus, etc), 
have recently been recognized as potential solutions for the aforementioned shortcomings of 3D noble metals.\cite{nl_2011_GPs,next10} This view is supported by theoretical calculations 
 which predict that graphene is able to sustain long-lived plasmonic excitations,\cite{low2016polaritons,nl_2011_GPs,GoncalvesPeres,AbajoACSP,Xiao2016} owing to its 
 remarkable optoelectronic properties.\cite{RMP81,GoncalvesPeres} At the same time, 
the electronic density of its charge-carriers can be dynamically tuned by means of electrostatic gating.\cite{NatNano,nphoton7,ACS7} The latter provides active control over 
the plasmon resonance (which is proportional to $\sqrt{E_F}$, where $E_F$ denotes the Fermi energy of graphene), and therefore may be conveniently varied on-demand. Yet, 
while the tunability issue was experimentally shown to be lifted by using graphene as a plasmonic medium,\cite{NatNano,nphoton7,ACS7} 
so far the realization of graphene plasmons\cite{GoncalvesPeres,AbajoACSP,Xiao2016} with long life-times remains elusive.\cite{Tassin:2012,Dastmalchi:2016} 
Such a fact is thought to be direct consequence of a combination of defects and disorder 
introduced during the growth and/or nanolithography stages, resulting in samples with hindered crystallinity and thereby lower mobilities. Nevertheless, 
there are reasons to stay optimistic about the promised low-loss graphene plasmonics. Nanofabrication techniques continue to improve, and recently 
highly confined graphene plasmons with reduced damping were experimentally realized using graphene encapsulated in hexagonal boron nitride (hBN).\cite{woessner2015}

Graphene plasmons have been observed in various configurations including graphene ribbons, disks or anti-dot arrays.\cite{NatNano,NJP14,nphoton7,ACS7,Zhu:2014,Wang:16,Goncalves:16,GoncalvesOptica} In addition, 
quasi-1D plasmons sustained at the edge of a semi-infinite graphene layer have been theoretically studied in Refs. \citenum{PhysRevB.84.085423,PhysRevB.85.235444}, 
based on previous works by Fetter on 2D electron liquids.\cite{PhysRevLett.54.1706,PhysRevB.33.3717}

\begin{figure}[t]
  \centering
    \includegraphics[width=0.475\textwidth]{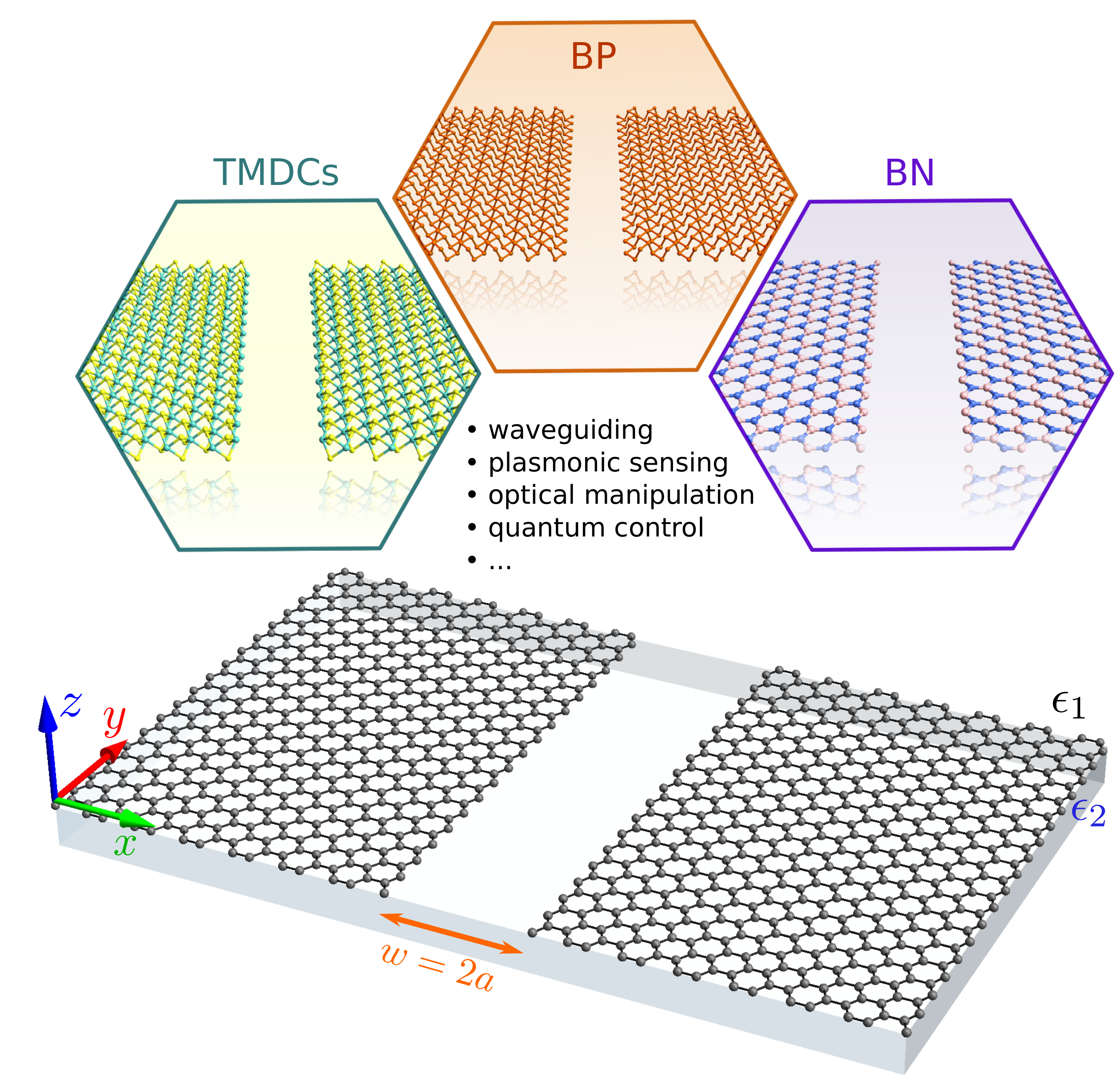}
      \caption{Illustration of the proposed 2D nano-slit using different 2D crystals, including doped graphene, transition metal dichalcogenides (TMDCs) 
      and black phosphorus (BP), and also hexagonal boron nitride (BN). While the latter does not support plasmon-polaritons, 
      it can sustain phonon-polaritons, and thus we include it here since their treatment is analogous. Although only monolayers are portrayed, 
      our framework is also applicable to their few-layer counterparts as long as a 2D conductivity can be attributed to them.}\label{fig:SYSscheme}
\end{figure}
In this work, we investigate plasmonic excitations arising in a nano-slit produced in a 2D material capable of sustaining plasmons, such as doped graphene, 2D TMDCs, 
or phosphorene (a black phosphorus monolayer); a scheme of the proposed structure(s) is depicted in Figure \ref{fig:SYSscheme}. 
The system can be fabricated by stripping a nanoribbon out of an otherwise homogeneous 2D layer, and therefore it is well within reach of current nanolithography techniques. As a more spectacular approach, we here mention the possibility for nanoparticle-assisted writing of nano-slits in graphene.\cite{Booth:2011} Finally, ultra-small nanowidths can be achieved by etching from a grain boundary.~\cite{Xie:2017} We further note that, in the spirit of Babinet's principle,\cite{BornWolf,PhysRevB.76.033407,ACS_bab} our system can be regarded as the complementarity structure to that of a 2D nanoribbon.
For small nano-slit gaps, the Coulomb interaction between edge plasmons in opposite 2D half-sheets gives rise to new hybrid plasmon modes of the compound system. 
Here, we demonstrate that the coupled plasmon resonances that emerge in such a structure split the dispersion relation of the bare edge plasmon 
into two new branches: one with lower energy and another with higher energy, akin to a bonding and an antibonding mode in molecular orbital theory. 
These findings are in line with plasmon hybridization models in metallic nanostructures, introduced in the 1960s~\cite{PhysRev.182.539} and further developed by Nordlander and co-workers.\cite{Prodan419} 
Our results, shown below, rely on a semi-analytical theory based on Green's functions and an orthogonal polynomial expansion\cite{PhysRevB.33.3717,GoncalvesPeres,wang2015,Goncalves:16}.  
This technique allows an accurate and reliable determination of the plasmonic spectrum, as well as the spatial distribution of the potential 
and corresponding electric field of the modes.~\cite{GoncalvesPeres,Goncalves:16}  Naturally, this information may then be used to estimate other related physical quantities, 
ranging from optical forces to the photonic density of states. Our treatment is comprehensive, encompassing both isotropic and anisotropic atomically thin layers.
Furthermore, 
we have also used commercial-grade full-wave numerical simulations (Lumerical and COMSOL)\cite{Lumerical,Comsol} to benchmark our semi-analytical results, to which we have observed an excellent agreement. 
A particularly attractive 
feature of our quasi-analytic model is that it entails a universal description of the plasmonic properties of nano-slits made in generic 2D crystals. This is because, as we detail below, 
the characteristics of the plasmon resonances depend uniquely on the system's geometry. %, which can be represented by single dimensionless parameter alone. 
Consequently, a single calculation is enough to uncover the plasmonic properties in 2D slits of all sizes. Such scale-invariance is 
courtesy of the electrostatic limit.~\cite{ACS6_Abajo,Jackson} % as in this regime an absolute length (which, in electrodynamics, is given by the free-space wavelength) is absent.\cite{ACS6_Abajo,Jackson} 
The consideration of the nonretarded regime here is fully justified since it provides an accurate description of plasmons in 2D materials as the corresponding plasmon wavevectors 
are typically much larger than $k_0=\omega/c$.\cite{GoncalvesPeres,AbajoACSP,Xiao2016} In passing, we remark that our notion of a 2D material 
could in principle also be extended to ultra-thin metallic films (which support a low-frequency plasmon).~\cite{Lee:11,Chen:2016,Raza_PRB88}

% ================================================== // ==================================================

% *************************************************************
% ::                       MAIN BODY                         ::
% *************************************************************
\section*{Results and Discussion}

{\bf Theoretical Background.} 
We consider an individual nano-slit carved out of an arbitrary 2D crystal, as sketched in Figure \ref{fig:SYSscheme}. We start by giving a brief account of 
the employed semi-analytic theory. The mathematical details are left as Supporting Information, where they are thoroughly dissected in an all-encompassing fashion.
In what follows, we work in the quasi-static limit since the effect of retardation is negligible when studying plasmons in 
atomically thin materials in typical experiments. Therefore, self-sustained plasmonic excitations in the system are governed by Poisson's equation, 
$\nabla^2 \Phi(\mathbf{r}) = -\varrho(\mathbf{r})/(\ep\ep_0)$. The translation invariance along the $y$-direction (cf. Figure \ref{fig:SYSscheme}) enables us to express 
the electrostatic potential as $\Phi(\mathbf{r})=\phi(x,z) e^{i k_y y}$ (and similarly for the charge-density, $\varrho(\mathbf{r})=\rho(x,z) e^{i k_y y}$), 
where a time-dependence of the form $e^{-i \omega t}$ is implicit hereafter. Hence, the Poisson's equation now takes the form
\begin{equation}
\left[ \frac{\partial^2}{\partial x^2} +\frac{\partial^2}{\partial z^2} - k_y^2 \right] \phi(x,z) = e\frac{n(x)}{\ep \ep_0}\delta(z) \ ,
 \label{def:Poisson_2D}
\end{equation}
where we have explicitly written the charge-density in terms of a delta function and a surface carrier density, i.e., $\rho(x,z) = -e n(x) \delta(z)$. The latter 
can then be expressed as a function of the in-plane electrostatic potential (see Supporting Information), which, together with the 
Green's function for a planar interface renders a self-consistent integro-differential equation for the induced 
(in-plane) potential, $\varphi(x)=\phi(x,z=0)$.~\cite{GoncalvesPeres} Such an equation does not possess an analytical solution, but one can still 
make further analytical progress by exploiting: 
\begin{enumerate*}[label={\roman*)}]
 \item the mirror symmetry of the considered nanostructure with respect to the plane bisecting the slit (defined by $x=0$); and, 
 \item by expanding the potential in one of the half-sheets using a basis containing Laguerre polynomials.~\cite{AS,Jackson,GoncalvesPeres} In particular, 
 taking $x>0$ without loss of generality, we write $\varphi^+ (x) = e^{-k_y(x-a)} \sum_{n=0}^{\infty} c_n L_n (2 k_y [x-a])$---please 
\end{enumerate*}
refer to Supporting Information for a detailed description. These steps are pivotal, and enable us to cast the above stated 
integro-differential equation into an elementary linear algebra eigenproblem, reading~\cite{noteEps}
\begin{equation}
 \frac{\ep_1 + \ep_2}{2 k_y} \frac{i \pi \omega \ep_0}{ \sigma_{\mathrm{2D}} (\omega)} c_n = \sum_{m=0}^{\infty} U_{nm}\ c_m \ , \label{eq:eigenproblem_G-slot}
\end{equation}
where the set of coefficients $\{c_n\}$ is determined by finding the eigenvectors of the matrix $\mathbf{U}$, whose matrix elements $U_{nm}$ are 
defined explicitly in the Supporting Information. On the other hand, the eigenvalues of $\mathbf{U}$, hereupon dubbed $\tilde{\lambda}$, uniquely 
determine the dispersion of the hybrid plasmon excitations guided along the 2D nano-slit, via
\begin{equation}
 \frac{\ep_1 + \ep_2}{2 k_y} \frac{i \pi \omega \ep_0}{\sigma_{\mathrm{2D}} (\omega) } = \tilde{\lambda}(\beta) \ . \label{res:DR_general}
\end{equation}
Clearly, Eq. (\ref{res:DR_general}) is completely general irrespective of the particular form of $\sigma_{\mathrm{2D}} (\omega)$. Thus, provided 
that one possesses an expression for the 2D conductivity, the plasmonic spectrum directly follows from the condition (\ref{res:DR_general}). 
We stress that, in its present form, Eqs. (\ref{eq:eigenproblem_G-slot}) and (\ref{res:DR_general}) assume that the surface conductivity is isotropic. 
We shall relax this assumption later in the paper, when considering the case of a nano-slit made from an anisotropic 2D material.

% ===============================================================================================================================================
%
\begin{figure*}[h]
  \centering
    \includegraphics[width=1.0\textwidth]{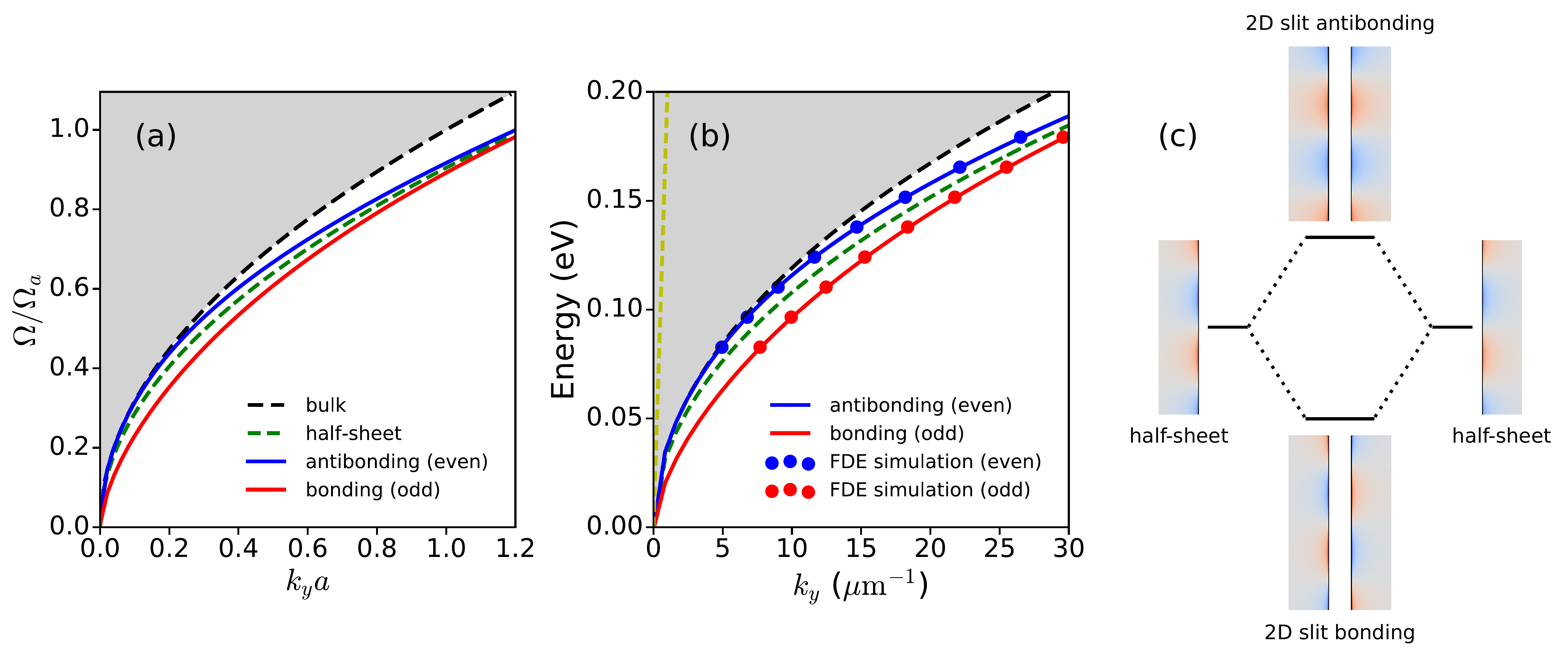}
      \caption{Dispersion relation of hybrid edge plasmons supported by a 2D nano-slit waveguide. The spectrum consist of two plasmonic bands, corresponding 
      to a bonding (odd) and an antibonding (even) mode originating from the interaction between edge plasmons in opposite 2D half-sheets. 
      (a) Spectrum of coupled plasmons in a 2D nano-slit, stemming from Eq. (\ref{eq:G-slot_dispersion}), 
      where $\Omega_a \equiv \Omega_\mathrm{bulk} (a^{-1})$. (b) Plasmon dispersion for a representative $w=2a=50\mathrm{nm}$ graphene nano-slit 
      calculated via Eq. (\ref{res:DR_general}), where we have employed Kubo's formula at $T=300$K for the conductivity of graphene 
      (other parameters are $E_F=0.5$ eV, $\Gamma=3.7$ meV, and $\ep_1=\ep_2=1$). The colored circles correspond to data points obtained from full-wave numerical 
      simulations based on FDE analysis (in frequency domain).~\cite{Lumerical}
      As per our quasi-analytical model, a number of $N_{\mathrm{max}} = 20$ terms were included for the truncation of the matrix $\mathbf{U}$---see Eq. (\ref{eq:eigenproblem_G-slot})---, and 
      shown to be sufficient in order to obtain a converged solution. (c) Plasmon hybridization scheme and calculated induced charge-densities~\cite{note_dp} of the plasmon modes, 
      obtained using our model. 
      }\label{fig:G-slot_spectrum}
\end{figure*}
{\bf Plasmon Dispersion.}
We now consider---for the sake of illustration and physical insight---a 2D nano-slit made of graphene with a frequency-dependent conductivity given 
by the Drude model, namely $\sigma_{\mathrm{2D}} (\omega) = \frac{i e^2 E_F}{\pi \hbar^2 \omega}$ (assuming negligible losses). Introducing this formula 
in Eq. (\ref{res:DR_general}) yields a closed-form expression for the dispersion of the graphene plasmon resonances in the system 
\begin{equation}
 \Omega (k_y) = \sqrt{ \frac{2}{\pi} \tilde{\lambda}(\beta) } \Omega_\mathrm{bulk} (k_y) \ , \label{eq:G-slot_dispersion}
\end{equation}
where $\Omega_\mathrm{bulk} (k_y) = \sqrt{ \frac{4 \alpha E_F \hbar c k_y}{\ep_1 + \ep_2} }$ is the nonretarded 
dispersion relation of plasmons propagating with wavevector $k_y$ in bulk (i.e., continuous) graphene \cite{GoncalvesPeres,AbajoACSP,Xiao2016} 
($\alpha \simeq 1/137$ denotes the fine-structure constant). 
If the atomically thin layer is instead a 2D conductor with parabolic dispersion, then expression (\ref{eq:G-slot_dispersion}) 
still holds but now with $\Omega_\mathrm{bulk} (k_y) = \sqrt{ \frac{4 \pi \alpha (\hbar c)^3 k_y}{\ep_1 + \ep_2} \frac{n_e}{m^{*} c^2} }$, 
where $n_e$ and $m^*$ are the carrier density and effective mass, respectively~\cite{2D_Drude}.
The reader should appreciate that, under its elegant and compact form, 
Eq. (\ref{eq:G-slot_dispersion}) entails a comprehensive description of the effect of the nano-slit's width in the coupling and 
subsequent hybridization of the modes sustained at opposite edges of the structure. Such information 
is contained in the eigenvalues $\tilde{\lambda}(\beta)$, which essentially depend on the dimensionless parameter $\beta=k_y a$, 
or, in other words, on the slit width to plasmon wavelength ratio. Therefore, 
the complete knowledge of the plasmonic spectrum can be fetched by diagonalizing the matrix $\mathbf{U}$ for a set 
of $\beta$-values, and then inserting the determined eigenvalues into Eq. (\ref{eq:G-slot_dispersion}). The outcome of such 
operation is shown in Figure \ref{fig:G-slot_spectrum}. 
As a guide to the eye, we have also included the spectrum of edge plasmons 
supported by an individual 2D half-sheet (green dashed line), as well as the plasmonic band of the corresponding bulk 2D plasmon (black dashed line). 
We stress that the results presented in Figure \ref{fig:G-slot_spectrum}a are valid for an arbitrary (isotropic) 2D crystal (with the appropriate choice of $\Omega_\mathrm{bulk}$, 
as discussed above). 
Notice that, as we have anticipated, 
the hybridization between the edge modes of the two half-planes results 
in a splitting of the unperturbed half-sheet edge plasmons into a pair of new hybrid modes. 
These arise from antisymmetric and symmetric hybridizations of the bare edge plasmons, giving rise to 
a bonding and an antibonding branch, lying below and above, respectively, the plasmon band of a single half-sheet---cf. Figure \ref{fig:G-slot_spectrum}. 
As the name suggests, in the case of the bonding mode the induced charge-density in opposite semi-infinite layers oscillate in anti-phase 
(odd symmetry), whereas for the antibonding mode such oscillations are in-phase (even symmetry). Naturally, these resonances emerge at 
different frequencies (for a fixed propagation constant), hence giving rise to the aforementioned energy splitting, as illustrated in Figure \ref{fig:G-slot_spectrum}c. 

It is apparent from Figure \ref{fig:G-slot_spectrum} that, at large wavevectors 
(in relation to the momentum-scale introduced by $a^{-1}$, i.e., for $k_y a \gg 1$),  
both resonances converge asymptotically to that of an edge plasmon in an individual half-plane. In this limit the 
Coulomb interaction between the neighboring edges falls off rapidly, and therefore the two semi-infinite 2D sheets decouple. 
Indeed, within this regime, the eigenvalue $\tilde{\lambda}(\beta \rightarrow \infty) \equiv \tilde{\lambda}^{(0)}$ 
becomes independent of $\beta=k_y a$. As a result, the dispersion relation of the modes of the 2D slit reduces to the plasmon 
of the unpatterned system 
multiplied by a proportionality constant, that is, $\Omega (k_y) = \mathrm{C^{te}} \cdot \Omega_\mathrm{bulk} (k_y)$. 
This behavior can be seen directly by inspecting the (analytical) integral kernel of the matrix elements $U_{nm}$, thanks to 
the amount of analytical progress performed here---check Supporting Information. This signifies that at large $k_y a$ the 2D nano-slit modes 
tend to become indistinguishable from that of an single half-sheet. 
Using our quasi-analytical theory we have obtained $\mathrm{C^{te}} = \sqrt{ 2 \tilde{\lambda}^{(0)} / \pi} = 0.905$ (see Supporting Information for more details) 
which is in outstanding agreement with the value of $\mathrm{C^{te}} = 0.906$ reported 
in the literature for edge plasmons supported by a 2D half-plane~\cite{volkov1988edge}.

It should also be stressed that Figure \ref{fig:G-slot_spectrum}a depicts the dispersion of the hybrid plasmon modes of the system in dimensionless units, thus 
enabling the determination of the plasmonic spectrum in 2D nano-slits of any width, and irrespective of the type of 
plasmon-supporting 2D material. %, as long as its 2D conductivity follows the Drude model. 
Necessarily, 
the exact positions of the resonances depend on the details of the model used for the 2D conductivity, but the general features outlined above 
should remain qualitatively the same. If the physics of the material cannot be captured by the Drude model, 
then one simply needs to take a step back and solve Eq. (\ref{res:DR_general}) directly, 
using an expression for the conductivity within the appropriate framework. 
In that context, Figure \ref{fig:G-slot_spectrum}b shows the solution of the condition (\ref{res:DR_general}) for a $50$nm-wide graphene 
nano-slit where the graphene has been modeled using Kubo's formula for the conductivity at finite temperature.~\cite{GoncalvesPeres} 
In the same figure, we compare the results of our semi-analytical theory against data obtained from 
numerical full-wave electrodynamic simulations using a commercially available finite-difference eigenmode (FDE) solver~\cite{Lumerical} (circles). The observed agreement between 
both techniques is quite remarkable. Such a fact, together with the ability to reproduce the $\beta \rightarrow \infty$ limit, 
demonstrates the ability of our quasi-analytical model to accurately describe plasmonic excitations in 2D nano-slits. All of this with the added advantage a semi-analytical method provides in portraying a clear and intuitive picture of the physics, without the necessity of relying on often time-consuming numerical simulations.

% ===============================================================================================================================================
{\bf Potential and Electric-Field Distributions.} 
The solution of the eigenproblem posed by Eq. (\ref{eq:eigenproblem_G-slot}) not only determines the plasmonic spectrum from 
the eigenvalues $\tilde{\lambda}(\beta)$, but also endow us the scalar potential within the 2D layer directly, by virtue of the eigenvectors of $\mathbf{U}$ (whose 
entries contain the set of expansion coefficients $\{ c_n \}$). 
\begin{figure}[h]
  \centering
    \includegraphics[width=0.475\textwidth]{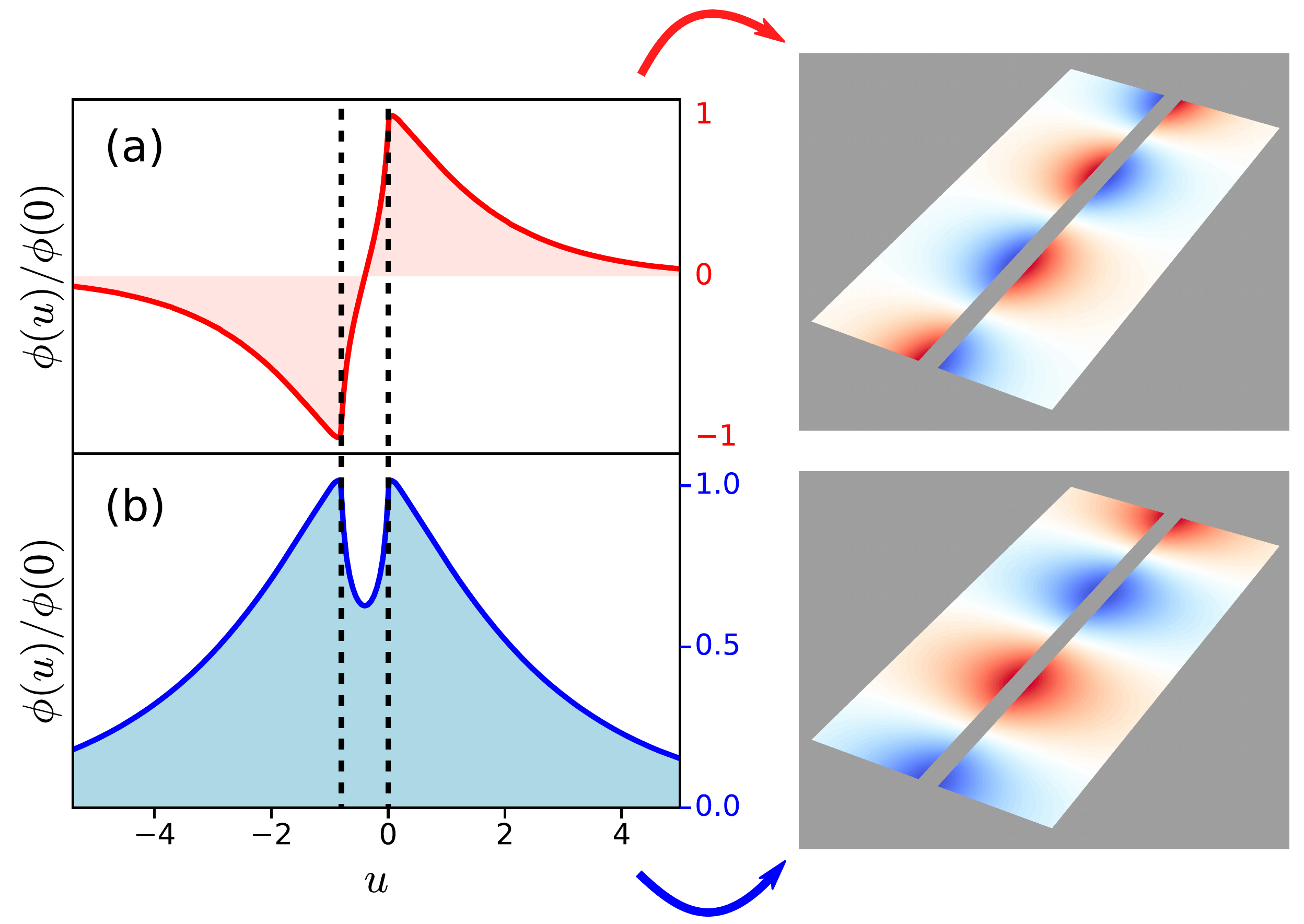}
      \caption{In-plane electrostatic potential along $u=k_y x - \beta$, i.e., $\phi(u,\zeta=0)$, 
      akin to the (a) bonding and (b) antibonding hybrid edge plasmons modes, for $\beta=k_y a = 0.4$. The corresponding panels at the right indicate 
      the electrostatic potential $\Phi(u,k_y y,0)$ evaluated within the 2D material. All plots are normalized to the respective maximums.
      }\label{fig:in-plane}
\end{figure}
From here, the electrostatic potential in the entire coordinate space follows from 
\begin{align}
  \phi (u,\zeta) &= -\frac{1}{2 \tilde{\lambda}(\beta)} \left\{ 
  \int_{0}^{\infty} dv \mathcal{K}_\eta (\beta;u,v,\zeta) \right.
  \left[ \frac{\partial^2 \varphi^+ (v)}{\partial v^2} - \varphi^+ (v) \right]  \nonumber\\
  &+ \left. \left. \frac{\partial \varphi^+ (v)}{\partial v}\right|_{v=0} \mathcal{K}_\eta (\beta;u,0,\zeta)
   \right\} \ , \label{eq:integro_diff_eq_G-slot_recXZ}
\end{align}
where we have introduced the dimensionless variables $u=k_y(x-a)$, $v=k_y(x'-a)$, and $\zeta=k_y z$ for numerical convenience and generality, 
while also defining $\mathcal{K}_\eta (\beta;u,v,\zeta) = K_0(\sqrt{(u-v)^2 + \zeta^2}) + \eta K_0(\sqrt{(u+v+2\beta)^2 + \zeta^2})$. 
Both the bonding and antibonding modes can be obtained in this way upon choosing the parameter $\eta$ 
appropriately ($\eta=1$ for the antibonding and $\eta=-1$ for the bonding), together with 
the corresponding $\tilde{\lambda}(\beta)$ and $c_n$'s. Figure \ref{fig:in-plane} illustrates the plasmon-induced 
in-plane potential, for $\beta=0.4$ (for other values the behavior is qualitatively similar). 
The corresponding resonant frequencies can be obtained from Figure \ref{fig:G-slot_spectrum} 
by reading the intersection of the vertical line $k_y a=0.4$ with the dispersion curves. 
From Figure \ref{fig:in-plane} it is clear that---as already noted above---the bonding (antibonding)
mode possesses a potential distribution which is odd (even) with respect to the plane bisecting the nano-slit, with 
the former exhibiting a nodal line in the middle of the gap.

Finally, the electric field induced by the plasmon oscillations in the system may be readily derived 
from Eq. (\ref{eq:integro_diff_eq_G-slot_recXZ}) by taking the gradient of the potential, 
$\mathbf{E}(\mathbf{r}) = - \nabla \Phi(\mathbf{r}) $, which gives (componentwise) %for each component of the electric field
\begin{subequations}
\begin{align}
 E_x (u,\Upsilon,\zeta) &= - k_y \cos \Upsilon\ \frac{\partial}{\partial u} \phi(u,\zeta)
 \ , \label{eq:Eu}\\
 E_y (u,\Upsilon,\zeta) &= k_y \sin \Upsilon\ \phi(u,\zeta)
  \ , \label{eq:EY}\\
 E_z (u,\Upsilon,\zeta) &= - k_y \cos \Upsilon\ \frac{\partial}{\partial \zeta} \phi(u,\zeta)
  \ , \label{eq:Ezeta}
\end{align}
\label{eq:E_field}%
\end{subequations}
where $\Upsilon = k_y y$. In possession of the previous expressions, the electric field 
originating from each of the plasmon resonances can be fully determined at any given point 
in space. 
\begin{figure*}[h]
  \centering
    \includegraphics[width=0.9\textwidth]{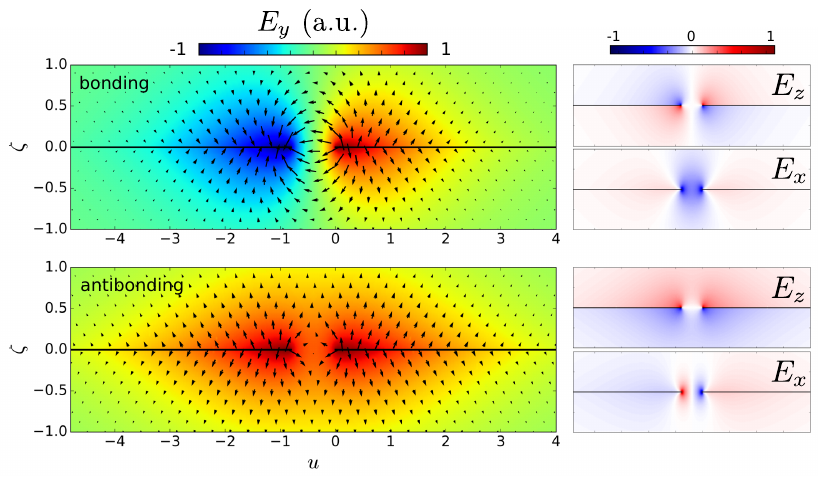}
      \caption{Density plots of the electric field distributions corresponding to the two hybrid edge plasmon modes of a 2D nano-slit. These are calculated 
      using the quasi-analytic theory described in the text---see Eqs. (\ref{eq:integro_diff_eq_G-slot_recXZ}) and (\ref{eq:E_field}). 
      We take $\beta=k_y a = 0.4$, and in the plots the cross-section of the slit is shown. The black solid line indicates the 2D plasmonic material.
      The three uppermost panels display the electric field akin to the bonding mode, whereas the ones at the bottom depict the 
      same quantities for the antibonding mode.
      The vectorial plots superimposed onto the two main panels illustrate the full 2D vector field, $\mathbf{E}(u,\zeta) = E_x(u,\zeta) \mathbf{u}_x + E_z(u,\zeta) \mathbf{u}_z$, 
      while the background, in rainbow colors, shows the $y$-component of the electric field (the component parallel to the edges of the nano-slit). The 
      length of the arrows is proportional to the norm of the electric-field vector at that point, $|\mathbf{E}(u,\zeta)|$, in logarithmic scale.
      Each one of the individual panels are normalized to their own maximum values, and the region depicted in the smaller panels has the same dimensions 
      as the main plots.
      }\label{fig:E_field}
\end{figure*}
The spatial distributions of all three components of the electric field in the $xz$- ($u\zeta$)-plane 
are depicted in Figure \ref{fig:E_field} (see the caption for details). 
This figure summarizes the main features of each of the hybrid edge plasmon modes in 2D nano-slits. 
The field components corresponding to the bonding mode are shown in the uppermost panels, while the ones 
corresponding to the antibonding mode are pictured in the lower part of the figure. In the main panels we 
have also superimposed the vector fields, represented by the arrows, associated with each plasmonic excitation. 
Naturally, the symmetries entailed in the electrostatic potential also emerge here. Interestingly, 
notice the similarity of the bonding eigenmode with that of an electric dipole oriented across the gap 
formed by the nano-slit. In contrast, the other resultant mode hybridization resembles two electric monopoles 
separated by a distance equal to the width of the gap separating the two half-sheets.~\cite{note_A} Finally,  
we note that in an experimental setting, a particular plasmonic resonance may appear as ``bright'' or ``dark'' 
under plane-wave illumination depending on the polarization of the incident field. Such criteria could be used, for instance, 
to select a single mode \emph{a priori}.
%

% ===============================================================================================================================================
{\bf Complementary 2D Structures: Slit and Ribbon.} 
The 2D nano-slit geometry considered here may also be perceived as 
the inverse (or complementary) structure to a 2D nanoribbon. Therefore, it 
is instructive to compare the character of the guided plasmon modes supported by these 
complementary 2D nanostructures (both assumed to be infinitely long, for the sake of simplicity). 

The ribbon system has been object of intensive study, 
specially inasmuch as graphene is concerned, where it has been shown that graphene ribbons sustain 
a set of discrete plasmon resonances.~\cite{NatNano,ACS6_Abajo,Rodrigo10072015,nphoton7,GoncalvesPeres,Zhu:13,Goncalves_PRB94} Each of such plasmon excitations 
are propagating along the ribbon direction, while resembling linear 
monopoles, dipoles, and higher-order multipoles in the transverse direction. 
These arise from the confinement of the plasmon wavenumber across the ribbon width, 
thus mimicking standing waves confined in a one-dimensional box. This contrasts 
with what we have found for the 2D nano-slit, and it simply reflects the absence 
of a finite size in the latter case (since the nano-slit is 
composed by two semi-infinite planes).%, or, in practice, when the size of the crystal is much larger than the gap width).

\begin{figure}[h]
  \centering
    \includegraphics[width=0.65\textwidth]{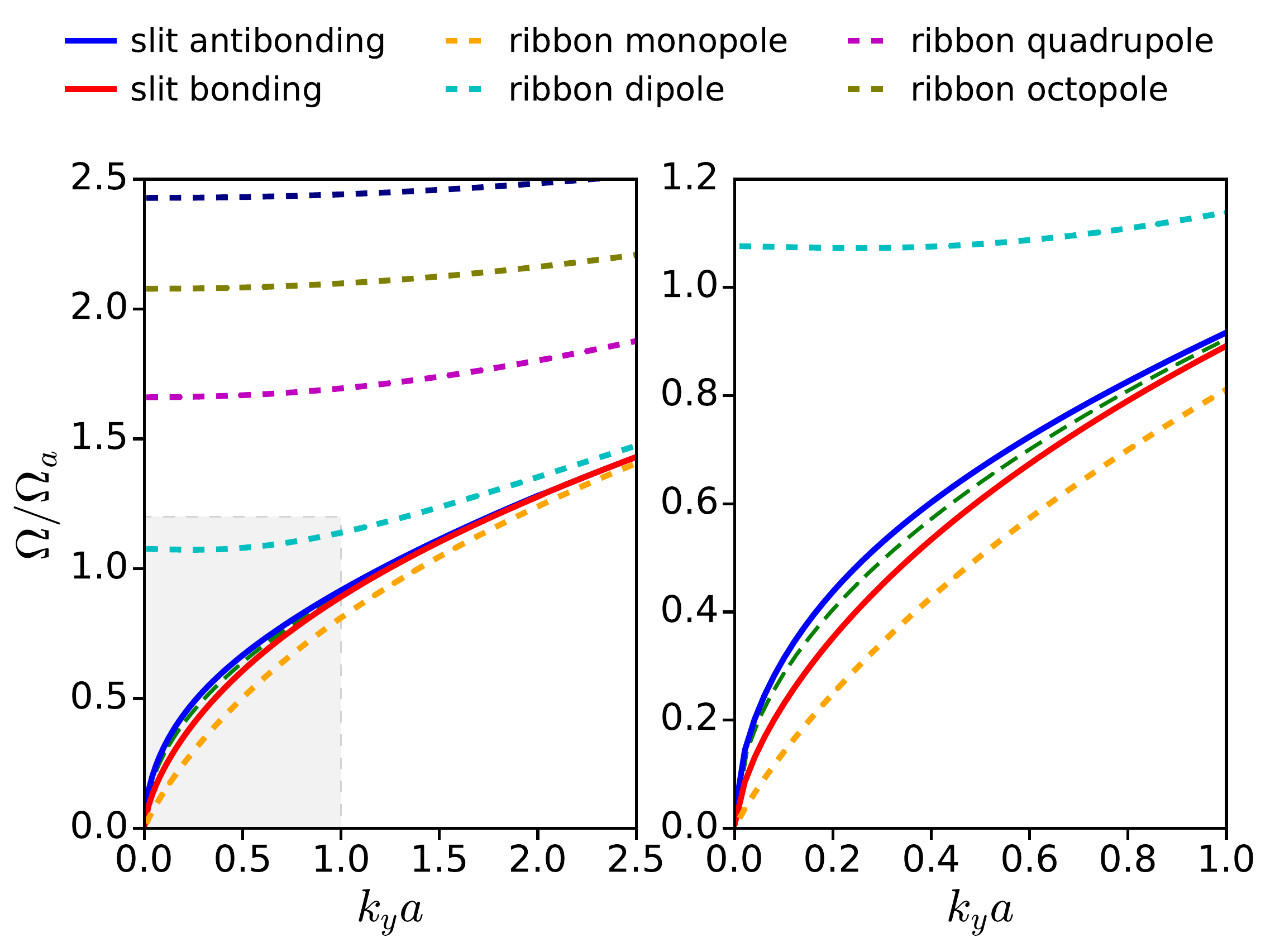}
      \caption{Plots of the dispersion relation of the hybrid edge modes in 2D nano-slits (solid lines) together 
      with the plasmonic spectrum of a nanoribbon (dashed lines). The latter is computed using a similar technique,  
      described elsewhere.~\cite{GoncalvesPeres} The panel at the right is a zoomed version of the 
      shaded area in the left panel. The thin green dashed line represents the edge plasmon in a single half-sheet, and serves as an eye-guide.
      }\label{fig:ribbon_vs_slit}
\end{figure}

Figure \ref{fig:ribbon_vs_slit} shows the dispersion of the 
hybrid edge plasmons of a 2D slit together with 
the plasmonic spectrum of a ribbon. We have plotted the 
dispersion relations in dimensionless units to facilitate 
the direct comparison between ribbons and slits of the 
same, but arbitrary, width $w=2a$. In what follows, we shall focus 
on the two lowest energy ribbon plasmons, which are the ones that admit a degree 
of similitude with respect to the modes of a slit. 
Indeed, analogous to the 
bonding and antibonding plasmon modes in a 2D slit, the monopole- and dipole-like ribbon 
resonances can also be understood as symmetric and antisymmetric hybridizations between the individual half-sheet plasmons 
living in opposite edges. This interpretation is further substantiated by the fact that both the dispersion of the slit hybrid modes and 
that of the aforementioned ribbon resonances approach the dispersion curve of the half-sheet edge plasmon at large $k_y a$---see Figure \ref{fig:ribbon_vs_slit}. 
In that limit, the plasmons in antipodal edges of the slit/ribbon effectively decouple and become indistinguishable from the edge plasmon of a 
single semi-infinite plane.

At this point, it is worth mentioning the differences and similarities between the 2D nano-slit and nanoribbon with their equivalent 3D versions, 
that is, the metal-insulator-metal (MIM) and the insulator-metal-insulator (IMI) structures, respectively. Simply put, we note that the former can be regarded as 
the lower dimensional form of the latter. In particular, like the 2D versions discussed here, the MIM and IMI waveguides also 
accommodate symmetric and antisymmetric hybridizations of the two surface plasmons of the individual insulator-metal interfaces. 
In addition, although perhaps not surprisingly, both the 2D slit and the MIM waveguide exhibit a low-frequency plasmon in which the 
induced charge-density oscillates in anti-phase, whereas for the high-frequency mode they oscillate in-phase; and vice-versa 
in the ribbon and IMI configurations. However, despite such lookalike features---save for the ones related to the distinct dimensionalities---there 
are some important differences. First, contrary to the ribbon, the IMI waveguide does not support standing wave like resonances within the thin-film. This 
can be attributed to the very effective screening within the 3D metal, as opposed to the reduced screening of a 2D ribbon (nevertheless, as soon as the 
IMI waveguide is patterned into a strip of a thin (but still 3D) film, such standing wave excitations promptly emerge). 
Lastly, the most interesting distinction between the 2D nano-slit/MIM and their corresponding inverse structures 
is in what the breaking of complementarity is concerned. It is well-known that, in traditional IMI/MIM waveguides, the 
dispersion of their plasmon modes is identical in the nonretarded limit.~\cite{Raza_PRB88} This property is direct consequence of the 
Babinet's principle of complementarity structures. The complementarity of IMI/MIM waveguides is immediately broken upon inclusion of 
retardation effects~\cite{Raza_PRB88}---see also Supporting Information. Incidentally, notice that the situation is remarkably different in the 2D case. Here, both the 2D slit and ribbon 
plasmon modes are essentially electrostatic in nature (i.e., retardation effects are negligible), 
but complementarity seems to be broken nonetheless---cf. Figure \ref{fig:ribbon_vs_slit}. This feature has no 
parallel in the corresponding 3D versions. Strictly speaking, Babinet's principle is only exactly valid for perfect conductors.~\cite{BornWolf} 
Thus, the breaking of complementarity may be possibly attributed to the significant extension of the induced electric field within the 2D material, 
which is comparatively larger than for a 3D metal owing to the skin effect (so that the latter is closer to a perfect conductor).

The apparent breakage of complementarity in mutually inverse 2D nanostructures is not only 
interesting from a fundamental viewpoint, but it may also be important in practice when building 
complex plasmonic structures with mixed solid-inverse geometries, say, for instance, a vertical stack 
of 2D nano-slit together with a nanoribbon. To the best of our knowledge, such in-depth studies in 2D plasmonic 
structures are still lacking in the literature (while, e.g., plasmons in graphene nanoholes and nanodisks have been previously 
studied~\cite{ACS7,Zhu:2014}, the breaking of complementarity was not investigated). We believe this to be an important topic, 
certainly deserving future investigations.

% ===============================================================================================================================================
{\bf Nano-slits of Anisotropic 2D Crystals.} 
In our discussion so far, we have implicitly assumed that the conductivity of the contemplated 2D material is isotropic. This is indeed 
the case for graphene and group-VI TMDCs such as MoS$_2$, WS$_2$, or MoSe$_2$. Nevertheless, that remains but a particular case of a 
broader picture. Notably, there has been a growing interest in the plasmonics of anisotropic 2D materials, either as a platform 
to enhance and tune their inherent optical birefringence~\cite{PhysRevB.95.201401,Yang17,TiS3} or in the context of hyperbolic nanophotonics.~\cite{PhysRevLett.116.066804,Gjerding:2017}
Examples of anisotropic 2D materials include black phosphorus (bP),~\cite{Low_bP,carvalho2016phosphorene} trichalcogenides like TiS$_3$,~\cite{ANGE,TiS3} and group-VII TMDCs (for instance, 
ReS$_2$).~\cite{ADFM} 
Among these, few-layer black phosphorus and its monolayer version---\emph{phosphorene}---have been the subject of remarkable attention from 
the nanophotonics community, owing both to its high carrier mobility and attractive optical properties.~\cite{carvalho2016phosphorene} 
For this reason, we will dedicate the next lines to the study of nano-slits made from anisotropic 2D crystals, subsequently focusing on the case of phosphorene.

The key aspect differentiating this case from the isotropic scenario considered above, lies in the fact that the surface conductivity of the atomically thin material 
is now a tensor. Conveniently, one can still profit from the work performed earlier in the isotropic setting by implementing a few basic modifications in order  
to contemplate the medium's anisotropy. We shall refrain ourselves from enumerating the mathematical details here, but they are provided in the Supporting Information. 
The corollary of such procedure is an eigenvalue problem resembling the one posed by Eq. (\ref{eq:eigenproblem_G-slot}), where now the matrix $\mathbf{U}$ is augmented by 
the addition of another matrix accounting for the anisotropy of the system. As before, the dispersion of the plasmon eigenmodes of the anisotropic 2D nano-slit are determined 
by the eigenvalues of the total matrix. Denoting these by $\xi$, we find that the spectrum of the guided modes in a doped phosphorene nano-slit follows
\begin{align}
 \Omega_{\mathrm{bP}} (k_y) = \sqrt{ \frac{2\ \xi(\beta)}{\pi}  }\ \Omega_\mathrm{2DEG} (k_y) 
\ , \label{res:G-slot_dispersion_bP}
\end{align}
where $\Omega_\mathrm{2DEG}(k_y)$ refers to the dispersion relation of plasmons in a homogeneous and isotropic 2DEG. All the details of the anisotropy 
and specificities of the 2D crystal are therefore contained in the ``anisotropic eigenvalues'', $\xi$. Indeed, the same formalism used for phosphorene can also be applied to other anisotropic 
materials, or even otherwise isotropic materials under the application of uniaxial strain, which effectively breaks the isotropy. 

\begin{figure}[t]
  \centering
    \includegraphics[width=0.65\textwidth]{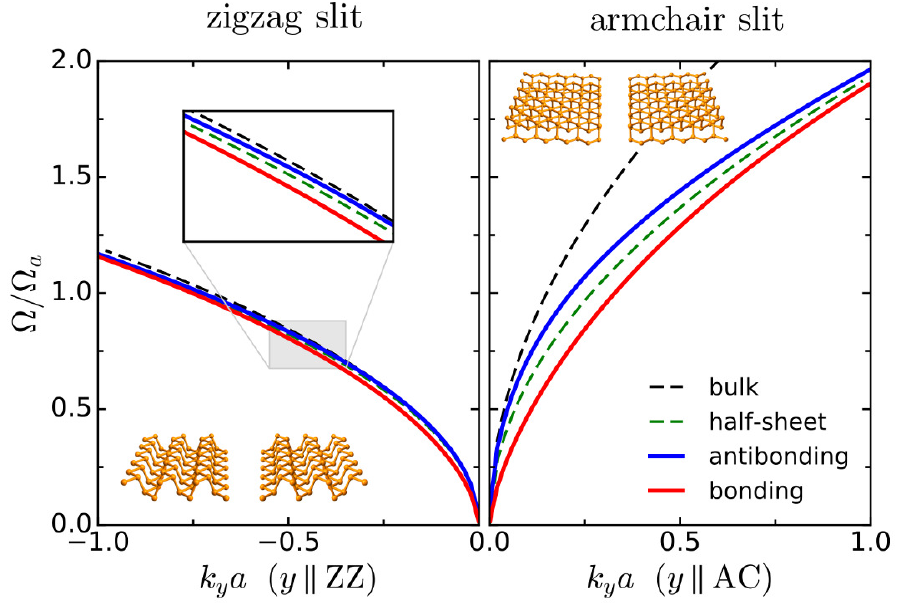}
      \caption{Plasmon eigenmode spectrum of the hybrid modes in a electron doped phosphorene nano-slit. Both the case of a 
      nano-slit patterned along the zigzag and the armchair directions is shown. We have assigned negative values of $k_y$ 
      to represent the hybrid plasmonic modes sustained at a zigzag slit, and positive values of $k_y$ to identify the eigenmodes 
      of an armchair nano-slit. In our calculations, we take the following parameters for the (anisotropic) electron effective masses: 
      $m_\mathrm{ZZ} = 0.7m_0$ and $m_\mathrm{AC} = 0.15m_0$,~\cite{Low_bP,carvalho2016phosphorene} for the electronic bands along 
      the zigzag and armchair directions, respectively. Additionally, here $\Omega_a = \Omega_\mathrm{2DEG} (k_y=a^{-1})$ (see Supporting Information).
      }\label{fig:bP}
\end{figure}

The low-energy bandstructure of monolayer bP can be approximated by that of an ordinary parabolic 2D semiconductor, whose conductivity 
can be constructed in terms of the carrier effective masses along the high-symmetry directions (zigzag and armchair).~\cite{PhysRevLett.116.066804,Low_bP,carvalho2016phosphorene} 
Figure \ref{fig:bP} shows the dispersion relation of anisotropic plasmons supported by a phosphorene nano-slit doped with electrons. The figure depicts two distinct situations: 
one where the edges of the slit run along the zigzag direction (left panel), and another in which the nano-slit is parallel to the armchair direction (right panel)---see 
also figure's insets. 
From Figure \ref{fig:bP}, it is clear that although the overall features already seen in the isotropic case remain---namely the existence of a bonding and an antibonding mode, 
respectively below and above the dispersion of the half-sheet plasmon---the behavior of the eigenmode spectra is strikingly different 
depending on the orientation of the 2D nano-slit with respect to the phosphorene's crystal axes. The dramatic contrast between the plasmon dispersion 
and hybridization in the two cases depicted in Figure \ref{fig:bP} reflects the strong anisotropy of black phosphorus 
(which in turn stems from its puckered honeycomb lattice). Hence, while the 
dispersion curves associated with the hybrid modes of a zigzag slit are barely indistinguishable from each other (and also from the half-sheet and bulk phosphorene plasmon) in the scale of Figure \ref{fig:bP}, 
the splitting between the bonding and antibonding modes of an armchair phosphorene nano-slit is substantial. This arises because, for instance in the latter case, 
the charge carrier effective mass is substantially lighter along the slit's edges (armchair) and it is heavier in the direction perpendicular to it. 
Naturally, the situation is reversed in the case of a zigzag nano-slit. 

As before, our quasi-analytic results for the anisotropic phosphorene nano-slit, embodied in Figure \ref{fig:bP}, were benchmarked and subsequently 
validated by rigorous electrodynamic simulations based on the finite-element method (a comparison is shown in the Supporting Information).~\cite{Comsol}

Therefore, in this manner, we have demonstrated that our semi-analytical model is extensible to nano-slits made from anisotropic 2D materials as well. 
Such a fact further enhances the applicability and versatility of our method to describe plasmonic excitations in a wide variety of 2D plasmonic materials, 
both with and without anisotropy. We therefore anticipate that our work may contribute to the development of anisotropic 2D plasmonics, since 
even the nanostructuring of challenging materials, such as phosphorene, is now well within reach of current experimental capabilities.~\cite{bP_ADL}

% ================================================== // ==================================================

% *************************************************************
% ::                      CONCLUSIONS                        ::
% *************************************************************
\section*{Conclusions and Outlook}

In conclusion, we have conducted a comprehensive theoretical description of the plasmon-mediated interaction 
and subsequent hybridization of plasmons in 2D nano-slits. Using a semi-analytical model, 
we have shown the emergence of two distinct waveguide-like eigenmodes of the compound system. 
These consist in a bonding and an antibonding plasmon resonance originating from the cross-talk between 
the two edge plasmons sustained at opposite margins of the slit. 
We fully characterize the dispersion relation of such modes, as well as the ensuing 
potential and electric field distributions. In order to gauge the accurateness of 
our technique, we also performed full-wave numerical simulations, 
which remarkably corroborated the results of the former. 
We note, however, 
that our semi-analytical framework paints a clearer and intuitive picture of the 
underlying physics. Furthermore, it also possesses the advantage of being universal, in the sense that---with a single calculation---it 
renders the plasmon dispersion for 2D slits of any width, and made from an arbitrary 
2D material. 
Going beyond the assumption of a nano-slit constructed from an isotropic 2D medium, the case of 
a black phosphorus (a strongly anisotropic 2D crystal) nano-slit was also investigated.

Moreover, the slit considered herein can be recognized as the inverse geometry of a ribbon. Therefore, 
we mutually compared our findings for the slit with the ones for the ribbon case, in the light of Babinet's principle 
of complementarity. Restricting our considerations to the plasmon dispersions, we have observed 
that Babinet's principle is only strictly satisfied in the asymptotic limit $k_y a \rightarrow \infty$. In opposition, 
in 3D IMI/MIM waveguides, complementarity is maintained within the nonretarded limit. This interesting subject 
remains largely unexplored in two dimensions, and further research would thus be valuable. 
Additionally, the combination of complementarity geometries could 
greatly augment the zoo of 2D plasmonic nanostructures.

In passing, let us underline that the theory described here is within a stone's throw to be exercised in many applications, 
for instance, to control the decay of quantum emitters (via Purcell enhancement) 
or for quantum information processing.~\cite{Asger_quantum,PhysRevLett.97.053002,Marquier:2017} 
The high degree of localization of the electric field near the nano-slit's edges also makes this 2D structure 
extremely well-suited for plasmonic sensing and SERS. Other applications include the optical manipulation and trapping 
of nanoparticles in the vicinity of the 2D nano-slit, due to the large gradients in the electric field intensity. 
Within the dipole approximation, and neglecting the scattering force, the near-field optical force exerted on the said 
particle is proportional to $\nabla |\mathbf{E}|^2$ (and to the real-part of the particle's polarizability).~\cite{Novotny} Hence, 
it is immediately apparent from Figure \ref{fig:E_field} that a dielectric particle will be subjected 
to significant near-field forces in the neighborhood of the 2D nano-slit (in particular, the trapping potential is deepest at 
electric field intensity maximums).

Summing up, we have demonstrated that 2D nano-slits are attractive candidates to exploit light-matter interactions at the nanometer 
scale using the continuously growing family of 2D materials, ranging from deep subwavelength waveguiding and 
sensing to quantum optical control. 

In closing, we emphasize that although here we have focused mostly on plasmon-polaritons, our theory can also be readily used 
to describe other polaritonic excitations in atomically thin crystals, such as 
phonon-polaritons in thin hBN slabs, or exciton-polaritons in TMDCs.\cite{Abajo_vdWreview,low2016polaritons} 
We believe that our investigation contributes with a new building block---a one-atom-thick nano-slit---to the 2D toolkit of 
hybrid plasmon resonances, thereby expanding our freedom and capabilities to design new tunable plasmonic systems based on flatland plasmonics.

% ================================================== // ==================================================
% -------------------------------------------------------------
% #                      Acknowledgments                      #
% -------------------------------------------------------------

\begin{acknowledgement}
P.A.D.G., S.X., and N.A.M. acknowledge the Danish National Research Foundation (Project DNRF103), through the Center for Nanostructured Graphene. 
N.A.M. is a VILLUM Investigator supported by the VILLUM FONDEN (grant No. 16498). 
N.M.R.P. acknowledges financial support from the European Commission (696656), and the Funda\c{c}\~ao Portuguesa para a Ci\^encia e a Tecnologia (UID/FIS/04650/2013). 
\end{acknowledgement}

\begin{suppinfo}
Supporting Information Available: In the Supporting Information we describe, in detail, the 
derivation of the semi-analytical model outlined here. Both cases, that of a 2D nano-slit made 
out of an isotropic 2D material and out of an anisotropic one, are addressed. 
We further comment on the effect of losses on the plasmon propagation and life-time. 
Finally, we briefly discuss the plasmon dispersion in complementarity 3D and 2D structures.
\end{suppinfo}

\bibliography{refs_G-slit}% Produces the bibliography via BibTeX.

\begin{mcitethebibliography}{85}
\bibitem[not()]{noteEps}
Notice that we have replaced $\ep$ by $(\epsilon_1+\epsilon_2)/2$ to correctly
  account for the generic case of two different dielectric media (as shown in
  Figure 1). We should stress that this no approximation, but simply a
  consequence of the electrostatic limit and a planar interface.\relax
\mciteBstWouldAddEndPunctfalse
\mciteSetBstMidEndSepPunct{\mcitedefaultmidpunct}
{}{\mcitedefaultseppunct}\relax
\EndOfBibitem
\bibitem[not()]{note_dp}
In the density plots exhibited in Figure 2c the the slit width has been
  slightly exaggerated for illustration purposes only. The correct scaling is,
  however, shown in Figure 3 when plotting the potential.\relax
\mciteBstWouldAddEndPunctfalse
\mciteSetBstMidEndSepPunct{\mcitedefaultmidpunct}
{}{\mcitedefaultseppunct}\relax
\EndOfBibitem
\bibitem[2D_()]{2D_Drude}
Recall that the Drude conductivity of a 2DEG is given by $\sigma_{\mathrm{2D}}
  (\omega) = \frac{i e^2 }{\omega} \frac{n_e}{m^{*}}$, in the absence of
  damping.\relax
\mciteBstWouldAddEndPunctfalse
\mciteSetBstMidEndSepPunct{\mcitedefaultmidpunct}
{}{\mcitedefaultseppunct}\relax
\EndOfBibitem
\bibitem[not()]{note_A}
A note of caution is in order here. Although at this point we have used the
  words 'dipole' and 'monopole', this is actually to be interpreted as a
  line---in the center of the slit and running parallel to its edges---of
  alternating dipoles or monopoles. For instance, ordinary dipoles possess
  cylindrical symmetry whereas in the current case the symmetry is with respect
  to the $z=0$ plane.\relax
\mciteBstWouldAddEndPunctfalse
\mciteSetBstMidEndSepPunct{\mcitedefaultmidpunct}
{}{\mcitedefaultseppunct}\relax
\EndOfBibitem
\end{mcitethebibliography}

\end{document}